# Application of the Covariant Projection finite elements in the E field formulation for wave guide analysis

V.S.Prasanna Rajan*, K.C.James Raju
School of Physics, University of Hyderabad, Hyderabad –500 046, India.

**Abstract :** The importance and the advantages of the covariant projection finite elements are highlighted . The covariant projection finite element is applied for two different cross sections of a rectangular wave guide. Its results are compared with the nodal cartesian component formulation validating the practical efficiency of the covariant projection formulation over the nodal cartesian formulation.

**Key words :** Covariant projection , cartesian formulation, spurious modes

**Introduction :** The finite element method has proved to be a versatile tool for the analysis of the electromagnetic field problems. In particular, the vector finite element method has been successful in dealing with problems concerning the computation of electromagnetic fields subject to the boundary conditions occurring by virtue of the physical shape of the geometry under consideration. However the finite element method had been troubled by the occurrence of the spurious modes in the solution of the resulting eigen value equation. The spurious modes are those non physical modes which are characterized by the non zero divergence of the magnetic field associated with that mode. Hence adequate care is necessary in the finite element analysis for modeling the zero divergence of the magnetic field to prevent the occurrence of spurious modes. Among the various approaches suggested for the prevention of spurious modes, one of the approach is to use the mixed order covariant projection elements in the series expansion of the unknown field and to enforce only the tangential continuity between the elements [1]. The covariant projections was initially introduced by Crowley, Silvester and Hurwitz for 3D vector field problems. With these elements, vector finite element methods can be implemented as directly as scalar finite element methods : boundary and interface conditions are easily imposed and spurious modes are removed without resorting to global penalty terms or constraints [2].

*Corresponding author : Email: vsprajan@yahoo.com, kcjrsprs@uohyd.ernet.in

**Theory :** The covariant projections of a vector field **E** are simply its components in a curvilinear coordinate system [2]. In terms of the components, the projections are represented as,

$$E_\xi = \mathbf{E} \cdot \vec{\xi} \qquad (1a)$$
$$E_\eta = \mathbf{E} \cdot \vec{\eta} \qquad (1b)$$
$$E_\upsilon = \mathbf{E} \cdot \vec{\upsilon} \qquad (1c)$$

The relationship between the covariant components of **E** to its cartesian components is given in the matrix form as,

$$\begin{bmatrix} E_\xi \\ E_\eta \\ E_\upsilon \end{bmatrix} = \begin{bmatrix} \frac{\partial \xi}{\partial x} & \frac{\partial \xi}{\partial y} & \frac{\partial \xi}{\partial z} \\ \frac{\partial \eta}{\partial x} & \frac{\partial \eta}{\partial y} & \frac{\partial \eta}{\partial z} \\ \frac{\partial \upsilon}{\partial x} & \frac{\partial \upsilon}{\partial y} & \frac{\partial \upsilon}{\partial z} \end{bmatrix}^{-1} \begin{bmatrix} E_x \\ E_y \\ E_z \end{bmatrix} \qquad (2)$$

Any point can be represented by nodal cartesian coordinates as (x,y,z). To introduce covariant projections, a new coordinate system with variables $(\xi,\eta,\upsilon)$ can be constructed so that only one variable changes along the side of the element. Unitary vectors

$$\mathbf{a}_\xi = \frac{\partial \mathbf{r}}{\partial \xi} \qquad \mathbf{a}_\eta = \frac{\partial \mathbf{r}}{\partial \eta} \qquad \mathbf{a}_\upsilon = \frac{\partial \mathbf{r}}{\partial \upsilon} \qquad (3)$$

can be defined which are tangential to the sides of the element. Corresponding reciprocal vectors can also be introduced from the base vectors given above. The unitary vectors, in general are neither of unit magnitude nor mutually orthogonal.

The reconstruction of $\mathbf{E}(\xi,\eta,\upsilon)$ from the unitary vectors given above is done as

$$\mathbf{E}' = E_\xi^e(\xi,\eta,\nu)\frac{\mathbf{a}_\eta \times \mathbf{a}_\nu}{v} + E_\eta^e(\xi,\eta,\nu)\frac{\mathbf{a}_\nu \times \mathbf{a}_\xi}{v} + E_\nu^e(\xi,\eta,\nu)\frac{\mathbf{a}_\eta \times \mathbf{a}_\xi}{v} \quad (4)$$

where $v = \mathbf{a}_\xi \cdot \mathbf{a}_\eta \times \mathbf{a}_\nu = \mathbf{a}_\eta \cdot \mathbf{a}_\nu \times \mathbf{a}_\xi = \mathbf{a}_\nu \cdot \mathbf{a}_\xi \times \mathbf{a}_\eta$ is the normalizing parameter.

The unknown functions $E_\xi(\xi,\eta,\nu), E_\eta(\xi,\eta,\nu), E_\nu(\xi,\eta,\nu)$ are given as

$$E_\xi(\xi,\eta,\nu) = \sum_{i=o}^{m}\sum_{j=0}^{n}\sum_{k=0}^{n} E_{\xi ijk} h_i(\xi) h_j(\eta) h_k(\nu) \quad (5a)$$

$$E_\eta(\xi,\eta,\nu) = \sum_{i=o}^{n}\sum_{j=0}^{m}\sum_{k=0}^{n} E_{\eta ijk} h_i(\xi) h_j(\eta) h_k(\nu) \quad (5b)$$

$$E_\nu(\xi,\eta,\nu) = \sum_{i=o}^{n}\sum_{j=0}^{n}\sum_{k=0}^{m} E_{\nu ijk} h_i(\xi) h_j(\eta) h_k(\nu) \quad (5c)$$

$m > 0, n > 0$, $E_{\xi ijk} E_{\eta ijk} E_{\nu ijk}$ are unknown coefficients.

$$h_0(\zeta) = 1 - \zeta, h_1(\zeta) = 1 + \zeta, h_i(\zeta) = (1-\zeta^2)\zeta^{i-2} \; i \geq 2 \quad (5d)$$

For 2-D, the coordinates are $(\xi,\eta)$ and $\mathbf{a}_z=1$, unit vector in z direction.

The corresponding covariant components are $E_\xi(\xi,\eta), E_\eta(\xi,\eta)$ which are given by

$$E_\xi = \sum_{i=0}^{m}\sum_{j=0}^{n} E_{\xi ij} h_i(u) h_j(v) \quad (6a)$$

$$E_\eta = \sum_{i=0}^{m}\sum_{j=0}^{n} E_{\eta ij} h_i(u) h_j(v) \quad (6b)$$

where the conditions on m,n an the h functions are the same as given in (5). The E vector is reconstructed as,

$$\mathbf{E} = E_\xi \left( \frac{\mathbf{a}_\eta \times \mathbf{a}_z}{|\mathbf{a}_\xi \times \mathbf{a}_\eta|} \right) + E_\eta \left( \frac{\mathbf{a}_z \times \mathbf{a}_\xi}{|\mathbf{a}_\xi \times \mathbf{a}_\eta|} \right) \qquad (7)$$

It is shown in [1] that in all the covariant projection expansions, the condition m<n and neither m nor n equal to zero and only the tangential continuity should be imposed for avoiding the spurious solutions. The limits of (ξ,η,υ) in all the covariant projection is from –1 to 1.

**E field formulation using the covariant projection :**

The **E** field finite element functional represented as a sum over elements is given by[3]

$$\Pi = \sum_e \Pi^e + \sum_{ob} \Pi^b + \sum_{ib} \Pi^b \qquad (8a)$$

where

$$\Pi^e = \frac{1}{2} \left[ \int_{\Omega^e} (\nabla \times \mathbf{E}^*)(\varepsilon^{-1} \nabla \times \mathbf{E}) d\tilde{\upsilon} - \int_{\Omega^e} (k_0^2 \mathbf{E}^* \cdot \mathbf{E}) d\tilde{\upsilon} + \int_{\Omega^e} (\mu^{-1}(\nabla \cdot \mathbf{E}^*)(\nabla \cdot \mathbf{E}) d\tilde{\upsilon} \right] \qquad (8b)$$

$$\Pi^b = \frac{1}{2} \oint_{\Gamma^e} (\mathbf{E}^* \times n) \cdot (\mu^{-1} \cdot \nabla \times \mathbf{E}) ds \qquad (8c)$$

and the super/sub script 'e ' refers to the individual element , 'ob' refers to the outer boundary segments and 'ib' refers to inter-element boundary contours and '* 'denotes complex conjugate of the field. The expression in (8a) indicates summation over elemental ,inter-element and outer boundary segment contributions. The expression in (8c) vanishes when the outer boundary of the waveguide is a good conductor , since the first term in the integrand vanishes on the surface of a good conductor. The integration over inter element boundary segments will also vanish as these integrals will appear twice but with opposite directions of integration and having the same integrand.

Hence,

$$\Pi = \sum_e \Pi^e \qquad (8d)$$

The curl and the divergence of **E** are given as follows.

$$\mathbf{E} = \begin{bmatrix} 0 & j\beta & \dfrac{\partial}{\partial \eta} \\ -j\beta & 0 & -\dfrac{\partial}{\partial \xi} \\ -\dfrac{\partial}{\partial \eta} & \dfrac{\partial}{\partial \xi} & 0 \end{bmatrix} \begin{bmatrix} E_\xi \\ E_\eta \\ E_\nu \end{bmatrix} \qquad (9a)$$

$$\nabla \cdot \mathbf{E} = \begin{bmatrix} \alpha \dfrac{\partial}{\partial \xi} & \alpha \dfrac{\partial}{\partial \eta} & \alpha \dfrac{\partial}{\partial \nu} \end{bmatrix} \begin{bmatrix} E_\xi \\ E_\eta \\ E_\nu \end{bmatrix} \qquad (9b)$$

$$\alpha = \dfrac{(1 - \cos(\text{angle between unitary vectors})}{\text{volume of the element}} \qquad (9c)$$

The **E** field variation with υ is assumed to have the form exp(-jβυ) and υ component of **E** is 90 degree out of phase as compared to the other components.

Substituting the matrix representations in the functional applying the condition of the vanishing of the variation of the functional with respect to the nodal electric field components, and assembling together after obtaining individual S and T matrices for each element leads to the matrix eigen equation of the form :

$$[S][E] - k_0^2 [T][E] = 0 \qquad (10)$$

which is to be solved for the eigen values $k_o^2 = \lambda$.

**Numerical Implementation :**

The covariant projection element is applied to rectangular wave guide with the following cross-sections :

a) Cross sectional sides parallel to the X and Y axes.
b) Cross sectional sides inclined to the X and Y axes.

The computer program written in TURBO BASIC for nodal cartesian formulation and covariant formulation for rectangular cross-section having its sides parallel and inclined to the cartesian axes were developed and the results are tabulated. The corresponding figures representing the two cases are given in Fig.(1) and Fig.(2) respectively.

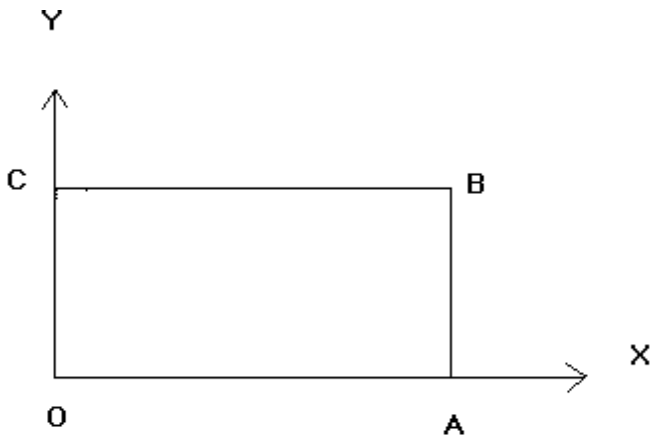

Fig 1. Cross sectional sides OA,BC parallel to the X axis and AB,OC parallel to the Y axis.

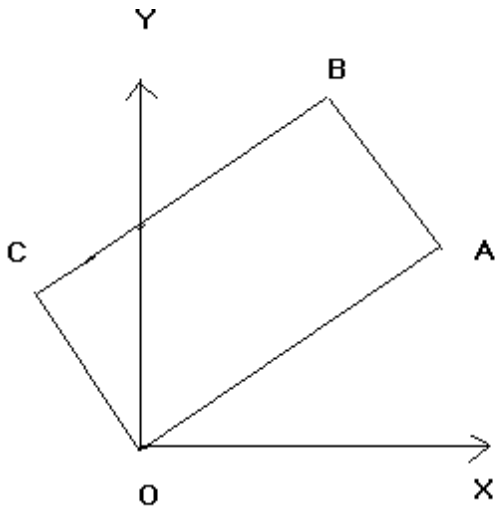

Fig 2. Cross-sectional sides OA and BC, inclined at an angle of 30* with respect to X axis. The sides OC and AB inclined at an angle of 30* with respect to Y axis.

The dimensions are taken as a=2.4 cm and b=1.2 cm. The cross-sectional geometry is divided into 72 rectangular elements having 4 nodes with each element. The total number of nodes are 91 and total number of field components are 273, being 3 components at each node. The program in TURBO BASIC was used to find the cut – off frequencies of different modes . The program was run for both orientations of wave guide cross-sections (ie rotated with 30 degree angle and without rotation). The results are given below in the form of a table.

**Cut off frequencies for wave guide with rectangular cross-section**

| Cross section orientation | E field component used | Cut –off for fundamental mode $k_o^2$ | 1st harmonic cut off $k_1^2$ | 1-comp type of condition | 2-comp type of condition | Time taken |
|---|---|---|---|---|---|---|
| Sides are II$^{el}$ to X and Y axes | Nodal cartesian | 1.7250 | 6.8935 | 148 | 0 | 15 min 26 secs |
| -do- | Nodal Covariant | 1.7250 | 6.8935 | 148 | 0 | 15 min 26 secs |
| Sides inclined (30 degs) | Nodal Cartesian | 1.7159 | 6.8931 | 68 | 8 | 19 min 25 secs |
| -do- | Nodal Covariant | 1.7250 | 6.8935 | 148 | 0 | 17 min 21 secs |

**Results and Discussion :** The above table shows that the cut off wave numbers of different modes obtained with covariant elements do not change due to rotation of the wave guide in cross-sectional plane. This is what is expected physically. Further, the absence of boundary conditions with two unknowns, less time is required for processing and elimination of spurious solutions are observed to be the advantages of covariant elements for the considered wave guides.

**Conclusion :** The application of the covariant projection elements for wave guide analysis is explained and its ease of implementation over the nodal cartesian formulation is validated by numerical implementation.

**Acknowledgement :** The authors thank the Council for scientific and Industrial Research (CSIR), New Delhi , India for providing the financial assistance in the form of Senior Research Fellowship to the first author in the research project sponsored by it.